# Influence of Gravity Waves on the Internal Rotation and Li Abundance of Solar-Type Stars

Corinne Charbonnel[1] & Suzanne Talon[2]

1. Observatoire de Genève, 51, ch. des Maillettes, 1290 Sauverny, Switzerland, and Laboratoire d'Astrophysique de Toulouse et Tarbes, CNRS UMR 5572, OMP, 14, Av.E.Belin, 31400 Toulouse, France

2. Département de Physique, Université de Montréal, Montréal PQ H3C 3J7, Canada

**The Sun's rotation profile and lithium content have been difficult to understand in the context of conventional models of stellar evolution. Classical hydrodynamic models predict that the solar interior must rotate highly differentially, in disagreement with observations. It has recently been shown that internal waves produced by convection in solar-type stars produce an asymmetric, shear layer oscillation, similar to Earth's quasi-biennial oscillation, that leads to efficient angular momentum redistribution from the core to the envelope. We present results of a model that successfully reproduces both the rotation profile and the surface abundance of lithium in solar-type stars of various ages.**

Rotation plays a crucial role in stellar evolution. Low-mass stars, including the Sun, are known to start their life with a large surface rotational velocity and then spin down with time because of a magnetically dominated stellar wind linked to their external convection zones. The interplay between the loss of angular momentum through a wind and its redistribution inside the star creates velocity gradients that induce mixing of elements. In the case of a fragile element such as lithium, which is destroyed by proton capture at a relatively low temperature (~2.5 million degrees) not too far below the convection envelope, surface depletion is thus expected. The lithium atmospheric abundance has been determined in many stars for which a fair estimate of the mass and age is feasible. These data allow an estimate of the extent, magnitude, and temporal evolution of chemical transport in the outermost stellar radiative regions, which may be linked directly to the instantaneous distribution of angular momentum in these objects. Furthermore, the quasi-flat seismic solar rotation profile (*1-2*) tells us that the characteristic time scale for the evolution of angular momentum has to be shorter than the age of the Sun.

Sophisticated stellar models that take into account hydrodynamic processes induced by rotation (i.e., meridional circulation and shear mixing) fail to reproduce the major observational constraints described above (*3-4*), although they are successful in reproducing the abundance anomalies and evolution characteristics of more massive stars (*5*). For solar-type stars with relatively extended convective envelopes that are strongly spun down by magnetic braking in their infancy, these models predict large rotation gradients within the interior, which are not consistent with helioseismology (Fig.1). This is due to the too-low efficiency of the invoked hydrodynamic instabilities in redistributing angular momentum.

Two mechanisms have been proposed to explain the near uniformity of the solar rotation profile. The first rests on the possible existence of a magnetic field in the radiation zone (*6-7*). The second invokes travelling internal gravity waves (hereafter IGWs) generated at the base of the convection envelope (*8-9*). For either of these solutions to be convincing, they must be tested with numerical models coupling these processes with rotational instabilities and should explain all the aspects of the problem, including the lithium evolution with time.

In conjunction with turbulent viscosity, IGWs in the Earth's atmosphere lead to a phenomenon known as the quasi-biennial oscillation (or QBO), which is an alternating pattern of eastward and westward mean zonal winds observed in the stratosphere close to the equator. This feature was long known by atmospheric scientists, but could be explained only when waves were added to numerical models of the atmosphere (*10-11*). The main reason for this is that momentum transport by waves does not have the same behavior as turbulence. First, waves take angular momentum in the region where they are produced and deposit it where they are damped, and they are thus able to accelerate regions very far from where they originate. Second, they tend to increase rather than reduce local shears.

In stars, IGWs are generated in the convection zone. At the convective/radiative interface, they produce a very narrow, double-peaked shear layer that oscillates on a time scale of a few years (*12-13*). This shear layer oscillation (hereafter SLO) is similar to the QBO. Talon, Kumar & Zahn (*14*) explained how this feature can lead to angular momentum extraction from the deep stellar interior when the outer convective zone is rotating slower than the interior. This is due to an asymmetry of the SLO, which preferentially damps the prograde waves that carry positive angular momentum. Low-degree, low-frequency waves then deposit preferentially negative angular momentum in the interior. However, these calculations were made in a static model and did not consider the issue of mixing of light elements. Later, we developed a formalism to take into account all aspects of IGWs in the computation of complete stellar evolution models (*15*). This model relies on WKB (Wentzel, Kramers, and Brillouin) approximation for the calculation of the local wave function and assumes that damping is caused by viscous turbulence and thermal diffusivity.

We used this formalism to compute self-consistent hydrodynamic models of evolving solar mass stars, including the transport of angular momentum by meridional circulation, shear turbulence (*16*), and IGWs. Elements evolve in this model by the same physical processes plus gravitational settling and nuclear reactions. All transport properties depend on the instantaneous rotation profile and distribution of elements. Initial surface rotation velocities typical of those observed for stars in very young open clusters were used. We applied braking by magnetic torquing at the stellar surface (*17*) so as to reach the observed rotation velocity at the age of the Hyades (*18*). We consider only IGWs produced by fluctuating Reynolds stresses (*19-20*). The total wave energy used for these calculations was $8.5 \times 10^{29}$ erg s$^{-1}$, that is 0.02% of the energy of the solar convection zone. Some uncertainty remains in the wave flux we use because of the eventual contribution of convective overshooting (*21-22*).

Figure 1 presents the evolution of the rotation profile for two cases: when angular momentum transport is due solely to meridional circulation and shear turbulence and when angular momentum deposition by IGWs is taken into account in conjunction with these hydrodynamic processes. In the former case, differential rotation remains large throughout the evolution, and its magnitude at the age of the Sun is excluded by helioseismology. When IGWs are considered, the low-degree waves penetrate all the way to the core and spin it down extremely efficiently at the very beginning of the evolution. This is related to the small (because it is proportional to $r^2$) amount of angular momentum in the core. Once the core has been spun down, the damping of retrograde waves, which carry the negative angular momentum, increases locally. Consequently a "slowliness" front forms and propagates in a wave-like way from the core to the surface. As further braking proceeds, a second front forms and propagates outward. The time scale for angular momentum extraction through differential wave filtering in a Sun-like star is of the order of a few

$10^7$ years (*8-9*). It adjusts itself so as to compensate for the flux of angular momentum that is lost through the stellar wind. This explains why front propagation is fast at the beginning and then slows down, just as the spin-down rate does.

Figure 2 shows the predicted evolution of the surface lithium abundance together with the data for solar mass stars in open clusters of various ages. In the case without IGWs, lithium depletion is always too strong. However, thanks to IGWs, the transport of elements and the resulting lithium depletion are considerably reduced because of the flattening of the internal rotation profile. Our calculations with IGWs fit the data quite well. The smallness of the observed dispersion in the lithium content is well explained even with a realistic and thus large range for the initial rotation velocity. This process is also self-regulating, and as such, our results do not depend qualitatively on the total wave flux used as long as it is large enough (that is, ~0.01% of the convective energy).

The presence of a dynamo magnetic field at the convective interface (termed tachocline) would not qualitatively change the results presented here. A strong magnetic field ($10^5$ G) may prevent very low frequency waves ($\omega < 0.1$ µHz for *l*=2, where $\omega$ is frequency and *l* is spherical harmonic degree) from propagating (*14-23*). The lowest frequency used for the calculations presented here is $\omega$=0.5 µHz. The low-degree waves that deposit angular momentum in the interior are thus not affected. Although the disappearance of high-degree, low-frequency waves could affect SLO dynamics, this has a negligible impact on filtering (*15*), which is dominated by the velocity difference on both sides of the SLO.

These results show in principle the ability of IGWs to efficiently extract angular momentum from the deep interior of solar-type stars on a very short time-scale and as such, nullify the argument made by Gough & McIntyre (*24*) about the "inevitability of a magnetic field" in the solar interior. Our hydrodynamic model, which uses the same free parameters to describe rotational mixing as those that successfully reproduce abundance anomalies in massive stars (*5*), successfully shapes both the rotation profile and the time evolution of the surface lithium abundance in these objects. In order to compare it to other models that rely on a fossil magnetic field (*25*), better helioseismic constraints are needed. The presence of a negative rotation gradient, for example, would strongly point toward wave transport. Our comprehensive picture should have implications for other difficult unsolved problems related to the transport of chemicals and angular momentum in low-mass stars. We think in particular of halo dwarf stars and the related cosmological problem of the primordial lithium, and of giants on the horizontal and asymptotic giant branches that exhibit unexplained abundance anomalies.

bottom, $h_\omega$ is the radial size of the largest eddy with frequency ω, $L$ is the size of an energy bearing eddy and $\tau_L \approx L/v$ is the convective time.

33. We would like to thank J.-O. Goussard, M. Grenon, G. Meynet, D. Pfenniger and D. Schaerer as well as anonymous referees and our editor for suggestions that improved the clarity of this text. Supported by Natural Sciences and Engineering Research of Canada (S.T.).

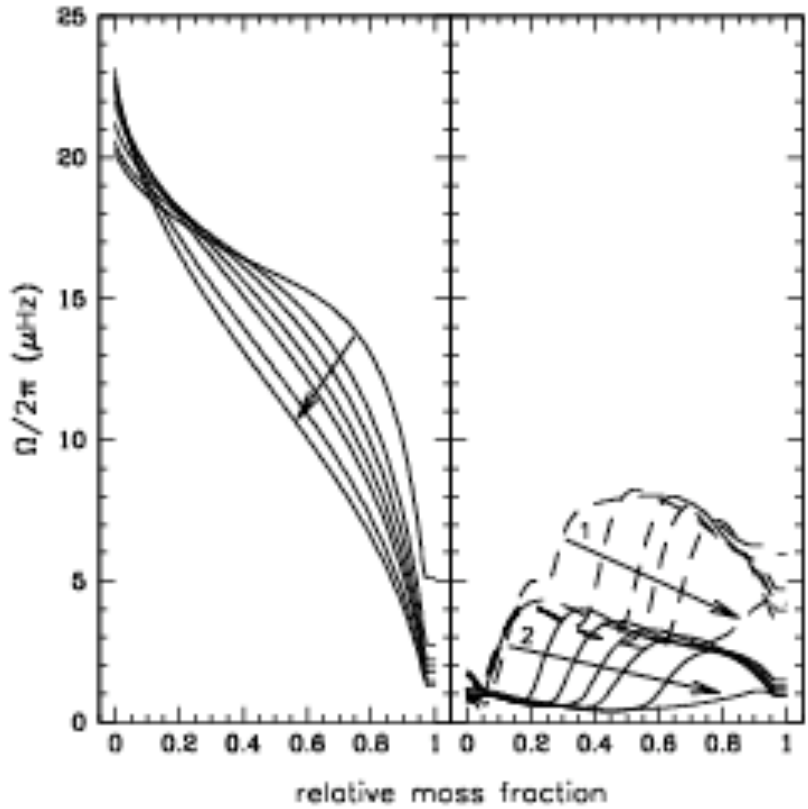

Fig. 1: Evolution of the interior rotation profile in a solar mass model with and without IGWs. The initial equatorial rotation velocity is 50 km s$^{-1}$, and identical surface magnetic braking is applied. (**Left**) Model without IGWs. Curves correspond to ages of 0.2, 0.5, 0.7, 1.0, 1.5, 3.0 and 4.6 billion years (Gy) and increase in the direction of the arrow. Differential rotation remains large at all times. (**Right**) When IGWs are included, low-degree waves penetrate all the way to the core and deposit their negative angular momentum in the whole radiative region. Because the core's angular momentum is minute, it is spun down very efficiently. In the so-created "slow" region, damping of retrograde waves increases, leading to the formation of a front, which propagates from the core to the surface. Curves showing propagation of the first front (labeled 1) correspond to ages of 0.2, 0.21, 0.22, 0.23, 0.25 and 0.27 Gy. Further spin down leads to the formation of a second

front (ages 0.5, 0.7, 1.0, 1.5, 3.0 and 4.6 Gy). The first front propagates faster than the second one due to stronger braking early in evolution. At the age of the Sun, the radiative region is rotating almost uniformly.

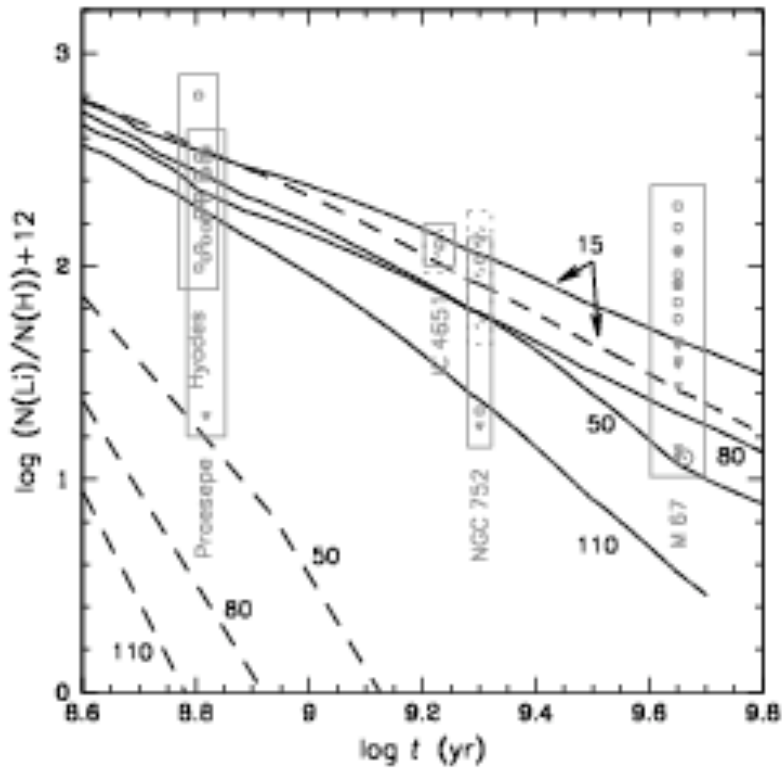

Fig. 2: Evolution of surface lithium abundance (N is the number abundance) with time for solar mass stars. The vertical extent of boxes shows the range of lithium values as observed in various galactic clusters (*26-32*) for stars with an effective temperature corresponding to that of the model ±100 K at the cluster age, plus a typical error in abundance determination. The horizontal extent corresponds to the age uncertainty. Circles indicate abundance determinations, and triangles denote upper limits for individual stars. The solar value is shown with the usual symbol ☉. Solid lines correspond to models including IGWs and dashed lines to models without IGWs. Initial velocities are shown on the figure (in km s$^{-1}$). In the cases

without IGWs, except for the slowest rotator, lithium depletion is too strong, by orders of magnitude, at all ages. When included, IGW, by changing the shape of the internal velocity gradients, lead to a decrease of the associated transport of chemicals. Lithium is then much less depleted and predictions account very well for the data. At all considered ages, the observed dispersion in atmospheric lithium is well explained in terms of initial velocity of each specific star. t, time in years.